\documentclass[twocolumn,preprintnumbers,amsmath,amssymb,prb]{revtex4}

\usepackage{xspace}
\usepackage{graphicx}
\usepackage{dcolumn}
\usepackage{bm}
\usepackage{tabularx}
\usepackage{amsmath}
\usepackage{amssymb}
\usepackage{color}

\begin{document}

\title{Elastoresistivity of heavily hole doped 122 iron pnictides superconductors \footnote{The edited version was published by Frontiers in Physics: https://doi.org/10.3389/fphy.2022.853717}}

\author{Xi\v{a}och\'{e}n H\'{o}ng$^{1,2,}\footnote{xhong@uni-wuppertal.de}$, Steffen Sykora$^{2,3,}\footnote{steffen.sykora@tu-dresden.de}$, Federico Caglieris$^{4,2,5,}$\footnote{federico.caglieris@spin.cnr.it}, Mahdi Behnami$^2$, Igor Morozov$^{2,6}$, Saicharan Aswartham$^2$, Vadim Grinenko$^{2,7,8}$, Kunihiro Kihou$^{9}$, Chul-Ho Lee$^{9}$, Bernd B\"{u}chner$^{2,8}$, and Christian Hess$^{1,2,}\footnote{c.hess@uni-wuppertal.de}$}

\affiliation{
$^1$Fakult\"{a}t f\"{u}r Mathematik und Naturwissenschaften, Bergische Universit\"{a}t Wuppertal, 42097 Wuppertal, Germany\\
$^2$Leibniz-Institute for Solid State and Materials Research (IFW-Dresden), 01069 Dresden, Germany\\
$^3$Institute for Theoretical Physics and W\"{u}rzburg-Dresden Cluster of Excellence $ct.qmat$, Technische Universit\"{a}t Dresden, 01069 Dresden, Germany\\
$^4$University of Genoa, Via Dodecaneso 33, 16146 Genoa, Italy\\
$^5$Consiglio Nazionale delle Ricerche (CNR)-SPIN, Corso Perrone 24, 16152 Genova, Italy\\
$^6$Department of Chemistry, Lomonosov Moscow State University, 119991 Moscow, Russia\\
$^7$Tsung-Dao Lee Institute, Shanghai Jiao Tong University, 200240 Shanghai, China\\
$^8$Institute of Solid State and Materials Physics and W\"{u}rzburg-Dresden Cluster of Excellence $ct.qmat$, Technische Universit\"{a}t Dresden, 01062 Dresden, Germany\\
$^9$National Institute of Advanced Industrial Science and Technology (AIST) , Tsukuba, Ibaraki 305-8568, Japan}

\date{\today}

\begin{abstract}
Nematicity in the heavily hole-doped iron pnictide superconductors remains controversial. Sizeable nematic fluctuations and even nematic orders far from a magnetic instability were declared in RbFe$_2$As$_2$ and its sister compounds. Here we report a systematic elastoresistance study of series of isovalent- and electron-doped KFe$_2$As$_2$ crystals. We found divergent elastoresistance upon cooling for all the crystals along their [110] direction. The amplitude of elastoresistivity diverges if K is substituted with larger ions or if the system is driven towards a Lifshitz transition. However, we conclude none of them necessarily indicates an independent nematic critical point. Instead, the increased nematicity can be associated with another electronic criticality. In particular, we propose a mechanism how elastoresistivity is enhanced at a Lifshitz transition.
\end{abstract}

\pacs{not needed}

\maketitle

\section{Introduction}

The "122" family, an abbreviation coined for BaFe$_2$As$_2$ and its substituted sister compounds, played an central role in the study of iron-based superconductors \cite{NSR}.
Those tetragonal ThCr$_2$Si$_2$-type structured compounds are blessed by the fact that sizeable single crystals with continuous tuneable doping can be prepared in a wide range, which is a crucial merit for systematic investigation of various ordered states.
Within the extended phase diagram of 122 compounds, the heavily hole doped region, including the end-members K/Rb/CsFe$_2$As$_2$ are of particular interests.
The superconducting transition temperature $T_c$ of Ba$_{1-x}$K$_x$Fe$_2$As$_2$ peaks at optimal doping $x=0.4$, and continuously decreases toward the overdoped (larger $x$) region. $T_c$ remains finite in the end-member $x=1$, while a change of the Fermi surface topology (Lifshitz transition) exists around $x=0.6 \sim 0.8$ \cite{XuN}. Although the $T_c$ vs. $x$ trend seems to be smooth across the Lifshitz transition, there are quite a lot of things happening here.
Vanishing electron pockets for $x > 0.8$ destroy the basis of the inter-pocket scattering induced $S\pm$ pairing symmetry which is generally believed as the feature of most iron based superconductors.
As a result, a change of the superconducting gap structure across the Lifshitz transition was observed experimentally \cite{XuN,Shin,CPL}.
Comparable pairing strength at the transition can foster a complex pairing state that breaks  time-reversal symmetry. Such exotic state was also demonstrated to exist around the Lifshitz transition \cite{VadimPRB,VadimNP}.
Very recently, a so-called "$Z_2$ metal state" above $T_c$ at the Lifshitz transition has been unveiled, with an astonishing feature of spontaneous Nernst effect \cite{VadimNP2}.

Electronic nematicity, a strongly correlated electronic state of electrons breaking the underlying rotational symmetry of its lattice but preserving translation symmetry, has been a wave of research in unconventional superconductors, particularly in the iron-based superconductors \cite{RPP,RF}.
Consistent experimental efforts have identified nematicity in all different iron-based superconductor families \cite{Chu,Hosoi,Hong,Kou,LiSL,Terashima}, accompanied by theoretical proposals of the intimate relationship between nematicity and superconducting pairing \cite{RF3,SL,DL,DLM}.
However, according to the previous background, we should not simply extend what is known in the under- and optimal-doped 122s to the very over-doped region. Whether nematicity exists and how it develops in this region needs independent censoring.

Indeed, nematicity in the heavily hole doped 122 turns out to be more elusive.
Heavily hole doped 122s stand out as a featured series because of their peculiar Fermi surface topology, isostructural phase transition and possible novel pairing symmetries \cite{Tafti,WangYQ,Ptok,AP}.
Nematically ordered states were suggested by nuclear magnetic resonance spectroscopy and scanning tunneling microscopy works on CsFe$_2$As$_2$ and RbFe$_2$As$_2$, and they were found to develop in different wave vectors than the underdoped 122s \cite{ChenXH,FengDL}.
Such nematic state far away from magnetic ordering challenges the prevailing idea that nematicity is some kind of vestigial order of magnetism \cite{RF2}.
An elastoresistance study further claims that a tantalizing isotropic (or XY-) nematicity is realized in the crossover region from dominating [100] nematicity in RbFe$_2$As$_2$ to [110] nematicity at the optimal doping \cite{Ishida}.
However, following works pointed out that elastoresistance in K/Rb/CsFe$_2$As$_2$ is actually contributed by the symmetric $A_{1g}$ channel, having little to do with the $B_{1g}$ or $B_{2g}$ channels which are related to nematicity  \cite{Wiecki1, Wiecki2}. Overall, the debate is still on for this topic.

In this \textbf{brief report}, we will not touch upon the nature of the possible nematicity of K/Rb/CsFe$_2$As$_2$. Instead, we confirm phenomenologically the existence of elastoresistance ($\chi^{er}$) in K/Rb/CsFe$_2$As$_2$ and find that its amplitude diverges exponentially with growing substituted ion size.
Besides, we present $\chi^{er}$ data on a series of Ba$_{1-x}$K$_x$Fe$_2$As$_2$ crystals crossing the Lifshitz transition. We observe, unexpectedly, a clear enhancement of $\chi^{er}$ from both sides of the Lifshitz point.
Although a presumptive nematic quantum critical point (QCP) might be of relevance, here we propose a rather more conventional explanation, based on a small Fermi pocket effect. Our results add a new novel phenomenon to the Lifshitz transition of the Ba$_{1-x}$K$_x$Fe$_2$As$_2$ system, and highlight another contributing factor of elastoresistance which has been almost ignored so far.

\section{EXPERIMENTAL DETAILS}

Single crystals of heavily hole-doped Ba$_{1-x}$K$_x$Fe$_2$As$_2$ were grown by the self-flux method \cite{Sai,Sai1,Sai2}. The actual doping level $x$ was determined by considering their structural parameters and $T_c$ values.
Elastoresistance measurements were performed as described in Ref.s \cite{Chu,Hosoi,Hong}. Thin stripe-shape samples were glued on the surface of pizeo actuators. The strain gauges were glued on the other side of the pizeo actuators to monitor the real strain generated.
In most cases, the samples were mounted to let the electric current flow along the polar direction of the pizeo actuators (R$_{xx}$), along which direction the strain was measured by the gauge. For one sample ($x$ = 0.68), an additional crystal was mounted with 90 degree rotated according to the polar direction (R$_{yy}$). More details are described in Section 3.3.
Sample resistance was collected with a combination of a high-precision current source and a nanovoltage meter. Due to the very large RRR ($R_{300K}/R_0$) values of the samples, special care was taken to avoid a temperature drift effect and the electric current was set in an alternating positive/negative manner in order to avoid artifact.

We point out that very noisy and irreproducible elastoresistance results can be acquired if DuPont$\circledR$ 4922N silver paint is used for making the contacts to the samples. On the other hand, samples contacted with EPO-TEK$\circledR$ H20E epoxy or directly tin-soldering gave nice and perfectly overlapping results. Given that DuPont$\circledR$ 4922N silver paint is widely used for transport measurements and indeed suitable for elastoresistance experiments of other materials (for example, the LaFe$_{1-x}$Co$_x$AsO series \cite{Hong}), we have no idea of why it does not work for heavily hole-doped Ba$_{1-x}$K$_x$Fe$_2$As$_2$ crystals. In this work, the presented data were collected by using the H20E epoxy. To avoid sample degradation, the epoxy was cured inside a Ar-glove box. A similar silver paint contact problem of K/Rb/CsFe$_2$As$_2$ crystals was also noticed by another group \cite{Wiecki2}.

\section{RESULTS AND DISCUSSIONS}

\subsection{Elastoresistance Measurement}

\begin{figure}[b]
\includegraphics[clip,width=0.49\textwidth]{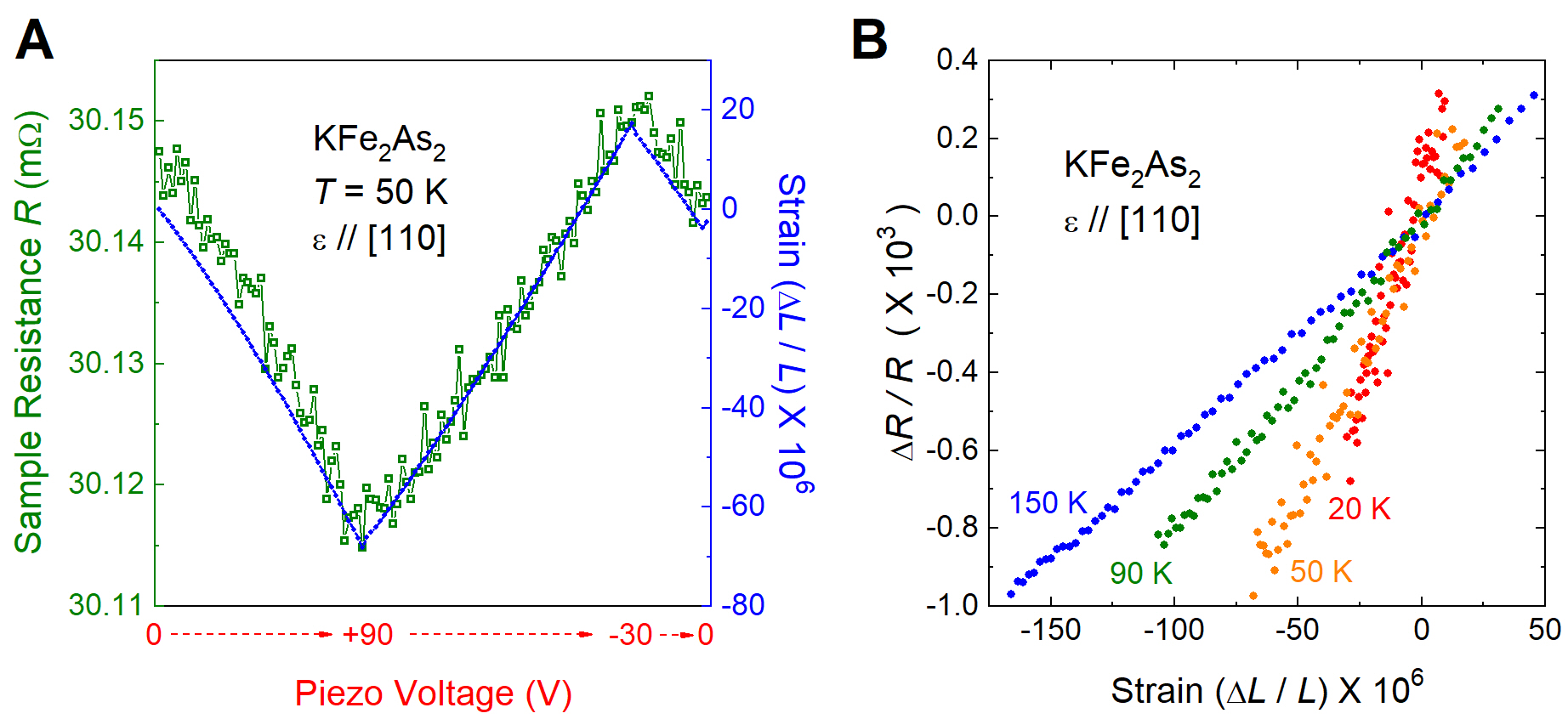}
\caption{Representative example of elastoresistance under strain for KFe$_2$As$_2$.
(A) Resistance and strain change according to the voltage applied across the piezo actuator at a fixed temperature $T$ = 50 K. The strain was applied along the [110] direction.
(B) The change of resistance $\Delta R/R$ as a function of strain $\Delta L/L$ at several temperatures.
}
\end{figure}

Elastoresistance measured along the [110] direction of KFe$_2$As$_2$ single crystal is exemplified in Fig. 1.
The sample resistance closely followed the strain change of the piezo actuator when the voltage across the piezo actuator is tuned.
As presented in Fig. 1B, the relationship between resistance change ($\Delta R/R$) and strain ($\Delta L/L$) is linear. This fact ensures that our experiments were performed in the small strain limit.
In such case, the elastoresistance $\chi^{er}$, defined as the ratio between $\Delta R/R$ and the strain, acts as a measurement of the nematic susceptibility \cite{Chu}.
It is worthwhile to notice that $\chi^{er}$ in KFe$_2$As$_2$ is positive (sample under tension yields higher resistance), consistent with the previous reports \cite{Ishida,Wiecki1}, and opposite to BaFe$_2$As$_2$ \cite{Chu}. Note that a sign reversal of elastoresistance was reported to occur at the underdoped region \cite{ECB}.

\subsection{Elastoresistance of K/Rb/CsFe$_2$As$_2$}

\begin{figure*}
\includegraphics[clip,width=0.95\textwidth]{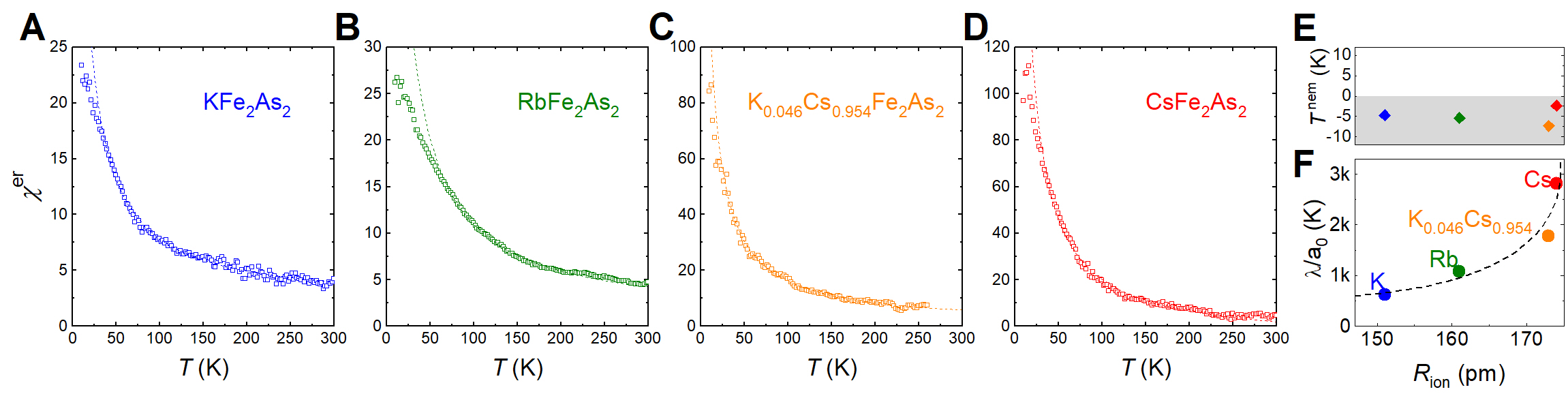}
\caption{
Temperature dependence of elastoresistance $\chi^{er}_{[110]}$ for (A) KFe$_2$As$_2$, (B) RbFe$_2$As$_2$, (C) K$_{0.046}$Cs$_{0.095}$Fe$_2$As$_2$, and (D) CsFe$_2$As$_2$. Solid lines are CW-fit of the data (see text). The fit parameters $T^{nem}$ and $\lambda/a_0$ are summarized in (E) and (F), respectively.
The dotted line in (F) is a guide to the eye.}
\end{figure*}

We start by showing our $\chi^{er}$($T$) data measured along the [110] direction ($\chi^{er}_{[110]}$) for a set of (K/Rb/Cs)Fe$_2$As$_2$ crystals. As clearly presented in Fig. 2, all the $\chi^{er}_{[110]}$($T$) curves follow a divergent behavior over the whole temperature range. A Curie-Weiss (CW) fit
\begin{equation}
\chi^{er} = \chi_{0} + \frac{\lambda/a_0}{T - T^{nem}},
\end{equation}
can nicely capture the data. A slight deviation can be discriminated at low temperatures which is typical for elastoresistance data, and is understood as a disorder effect \cite{Kou}. Note that the amplitude of the elastoresistance grows substantially from KFe$_2$As$_2$ to CsFe$_2$As$_2$, nearly 5-fold at 30 K.
The extracted parameters from the CW fit is shown in Fig. 2E and 2F. While the amplitude term shows a diverging trend, the $T^{nem}$ of all four samples are of a very small negative value, which practically remains unchanged if experimental and fit uncertainties are taken into account. That is at odds with a possible nematic criticality in this isovalent-doping direction.
The enhanced $\chi^{er}_{[110]}$ might be a result of a presumptive QCP of unknown kind or a coherence-incoherence crossover \cite{QCP1,QCP2,crossover}. These can not be discriminated by our technique, and thus are beyond the scope of this report.

\subsection{Elastoresistance of overdoped Ba$_{1-x}$K$_x$Fe$_2$As$_2$}

\begin{figure}
\includegraphics[clip,width=0.49\textwidth]{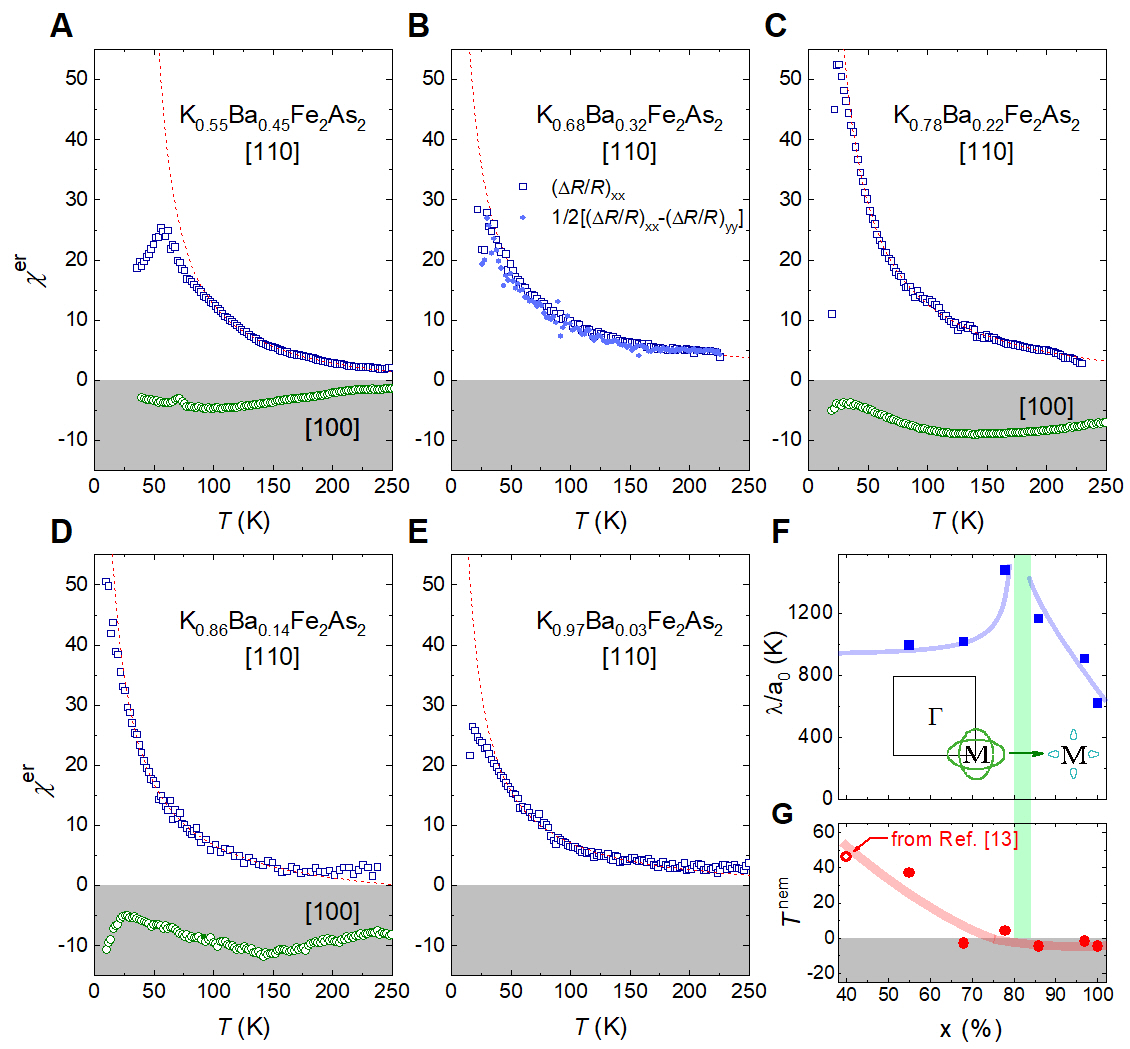}
\caption{Doping evolution of the elastoresistance in overdoped Ba$_{1-x}$K$_x$Fe$_2$As$_2$.
$\chi^{er}$ measured along the [110] direction is presented in the upper panels for Ba$_{1-x}$K$_x$Fe$_2$As$_2$ single crystals with (A) $x = 0.55$, (B) $x = 0.68$, (C) $x = 0.78$, (D) $x = 0.86$, and (E) $x = 0.97$. The red dashed lines are CW-fit to the data.
$\chi^{er}$ was also measured along the [100] direction for three of the samples. The data is presented in the lower panels of (A), (C) and (D).
In panel (B),$\chi^{er}$ extracted by using both $(\Delta R/R)_{xx}$ and $(\Delta R/R)_{yy}$ (filled light blue circles) and $(\Delta R/R)_{xx}$ (open blue squares) show indistinguishable results.
A doping dependence of the fit parameters are displayed in (F) $\lambda/a_0$ and (G) $T^{nem}$ of the [110] $\chi^{er}_{[110]}$ data. The thick lines are guide to the eye. A Lifshitz transition around $x=0.8$ is highlighted by the green bar. Hole pockets around the $M$ point of the Brillouin zone transform into electron lobes across the Lifshitz point \cite{XuN}. $T^{ nem}$ at $x=0.4$ is extracted from Ref. \cite{Kou}.}
\end{figure}

Next, we present a set of $\chi^{er}$($T$) data of five overdoped Ba$_{1-x}$K$_x$Fe$_2$As$_2$ ($0.55 \leq x \leq$ 1) across the Lifshitz point.
The elastoresistance, measured only for the $R_{xx}$ direction, as has been performed regularly in many reports \cite{Chu,Hong,LiSL}, has been argued to be inconclusive for the end members (K/Rb/Cs)Fe$_2$As$_2$, as a result of dominating $A_{1g}$ contribution, instead of a $B_{2g}$ (or $B_{1g}$) component which is related to nematicity \cite{Wiecki1,Wiecki2}. However, such complexity are ruled out by taking $R_{yy}$ also into account for calculating $\chi^{er}$($T$) for one representative example $x =0.68$ (Fig. 3B). The $\chi^{er}$($T$) curves calculated by the two different ways match nicely.

After checking the potential $A_{1g}$ contribution to $\chi^{er}$ for a doping level close to the Lifshitz transition, we turn now to the data itself.
As shown in Fig. 3, the $\chi^{er}_{[110]}$($T$) curves of Ba$_{1-x}$K$_x$Fe$_2$As$_2$ also follow a CW like feature.
One can see a clear dip at around 50 K in Fig. 3A for the $x=0.55$ sample. In some reports \cite{Ishida}, such feature was taken as a signal for a nematic order.
Since no other ordering transition (structural, magnetic, and so on) has been ever reported in this doping range, we refrain from claiming incipient nematic order solely based on such feature. This might equally well explained by different origins. However, we also can not exclude its possibility.

On the other hand, we measured $\chi^{er}$ along the [100] direction ($\chi^{er}_{[100]}$) for several samples. They are presented in the shaded panels of Fig. 3. All of them are negative, small in amplitude, and none of them shows a CW-like feature.
Such observation is incompatible with the existence of $B_{1g}$ nematicity in heavily doped Ba$_{1-x}$K$_x$Fe$_2$As$_2$ series, which is in sharp contrast to what is reported for the closely related Ba$_{1-x}$Rb$_x$Fe$_2$As$_2$ series \cite{Ishida,B1g1,B1g2}. As a result, the so-called XY-nematicity is clearly ruled out in the Ba$_{1-x}$K$_x$Fe$_2$As$_2$ series.

One remarkable feature, however, can be safely concluded, namely that the amplitude of $\chi^{er}_{[110]}$ has a clear tendency to peak around $x = 0.8$, close to the Lifshitz transition. This becomes more clear in Fig. 3F, where the CW fit parameters are plotted against the doping level.
The question is why $\chi^{er}_{[110]}$ is increased at the Lifshitz transition? A nematic QCP is a potential explanation. However, as Fig. 3G shows, $T^{nem}$ drops from $\sim 45$~K of the $x=0.4$ (optimal doped) to $\sim 0$~K at the Lifshitz transition. Further doping does not drive $T^{nem}$ to the more negative side within our experimental resolution. This is not a typical QCP behavior. Besides, since $T_c$ across the Lifshitz transition is quite smooth, it seems not boosted by pertinent potential nematic fluctuations. Furthermore, three-point bending experiments did not see any anomaly in this doping range \cite{3pb}. All these facts seem to be incompatible with the more understood nematic QCP in the electron doped side \cite{RF4}. Hence, if it is really a nematic QCP, novel mechanisms need to be invoked.
This motivated us to seek for alternative explanations for enhanced $\chi^{er}$ in heavily doped Ba$_{1-x}$K$_x$Fe$_2$As$_2$. Below, we propose a conventional argument based on the small pocket effect, exempting from invoking a QCP to exist at the Lifshitz transition.

\subsection{Theory for enhanced elastoresistance at the Lifshitz transition}

\begin{figure}[b]
\includegraphics[clip,width=0.49\textwidth]{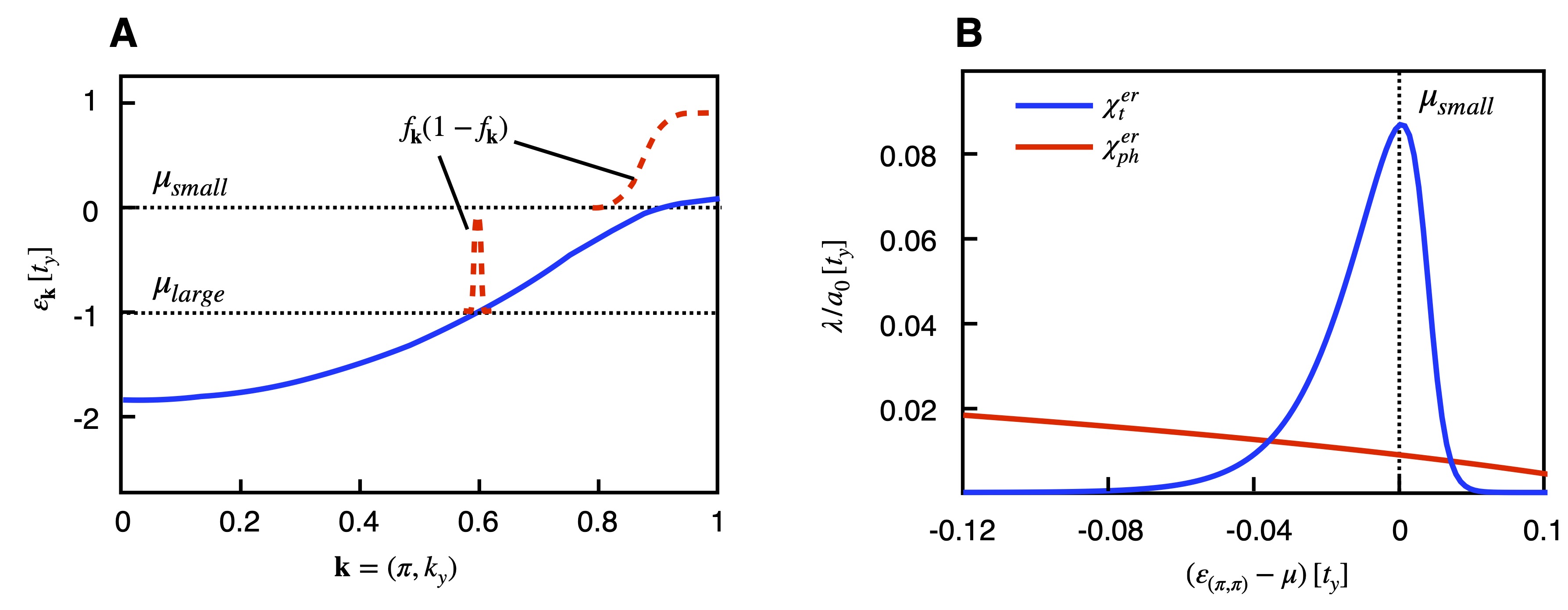}
\caption{Theoretical consideration of the elastoresisitivity. {\bf A}  Dispersion of the hole-like band leading to a very small hole pocket around the point $(\pi,\pi)$ when the chemical potential $\mu_{small}$ (upper dotted line) is placed near the Lifshitz point. The range of momentum vectors contributing to the elastoresisitivity arising from the Fermi function is schematically shown by the dashed lines. A much larger hole pocket is indicated by a lower value of the chemical potential $\mu_{large}$ (lower dotted line). {\bf B}  Calculated parts of the elastoresistivity according to Eq. \eqref{elastres_parts}. Around the Lifshitz point $\mu_{small}$ the first order part $\chi^{er}_t$ dominates the second order part which is, however, the most important contribution for $\mu$ away from the Lifshitz point.}
\end{figure}

To study the effect of a Lifshitz transition to the elastoresistance we have calculated this quantity based on a minimal model of iron-based superconductors \cite{Wuttke} with a very small Fermi surface. The corresponding dispersion which was used is shown in Fig. 4A for the normal state along a cut $(\pi , k_y)$.
We considered the two orbital model in Ref. \cite{Wuttke} with the same hopping matrix elements but having set the nematic interaction equal to a very small value. Thus, the nematic interaction accounts here only for the temperature dependence of the susceptibility according to a Curie Weiss law. Moreover, we introduced a very small lattice distortion in $x$ direction which is coupled to the electron system. Using first order perturbation theory with respect to this coupling (linear response) we then calculated the elastoresistivity.
We have considered two different cases of the coupling  between distortion and electrons (strength $g$). (i) The conventional coupling to the local electron density (electron-phonon coupling) where we denote the corresponding response with $\chi_{ph}$. (ii) A direct coupling of the distortion to the hopping matrix element $t_x$ in $x$ direction. The corresponding response is denoted by $\chi_{t}$.
\begin{equation}
\begin{split}
\chi_t &\propto g \lim\limits_{\Delta t_x\to 0} \frac{\sigma^{-1}_{xx}(t_x+\Delta t_x)-\sigma^{-1}_{yy}(t_x+\Delta t_x)}{\Delta t_x} \\
 \chi_{ph} &\propto \frac{1}{N} \sum_{\bf k} \left( \frac{g}{\omega + \varepsilon_{\bf k} - \varepsilon_{{\bf k}+{\bf q}}} \right)^2 \frac{f_{\bf k} - f_{{\bf k}+{\bf q}}}{\varepsilon_{{\bf k}+{\bf q}} - \varepsilon_{\bf k}}.
\end{split}
\end{equation}
Here $t_x$ is the hopping matrix element in $x$ direction and $\sigma_{xx},\sigma_{yy}$ are the conductivities in $x$ and $y$ directions \cite{Wuttke}. The phonon energy $\omega$ in $\chi^{er}_{ph}$ is in general renormalized by the coupling to the electrons and is becoming soft for a particular mode if the system is near a structural phase transition\cite{Sykora2006,Sykora2020}. The electron dispersion $\varepsilon_{\bf k}$ considered here is shown in Fig. 4A. The function  $f_{\bf k}$ is the Fermi distribution with respect to $\varepsilon_{\bf k}$. Thus,
for most systems investigated, this term dominates $\chi^{er}$. Note that how magnetic fluctuations impact on the phonons and the nematicity has been investigated \cite{RF, RF3}.

Fig. 4B shows the two parts $\chi^{er}_t$ and $ \chi^{er}_{ph}$ which were calculated separately as a function of the chemical potential $\mu$ to simulate different doping values. In order to compare with Fig. 3F, we extracted the temperature behavior according to Eq. 2 and plotted the calculated value $\lambda/a_0$ in energy units of $t_y$.   It is seen that the first order part  $\chi^{er}_t$ dominates over the second order part only in the narrow range of $\mu$ where the Fermi surface around $(\pi,\pi)$ becomes very small. Thus, only when the system has very small Fermi surfaces, as the case of Ba$_{1-x}$K$_x$Fe$_2$As$_2$ at the Lifshitz transition, the  term $\chi^{er}_t$ becomes important. However, it is also clearly seen that if the chemical potential is chosen away from the Lifshitz point corresponding to a proper doping the second order part $ \chi^{er}_{ph}$ is mostly important as expected.

The enhancement of $\chi^{er}_t$ in the presence of a very small Fermi surface can be explained by the existence of low energy excitations in a relatively wide range of momentum vectors. Since the conductivities are proportional to Fermi distribution functions $f_{\bf k}$ as follows,
\begin{equation}
\label{elastres_parts}
\sigma_{ii} \propto \sum_{\bf k} \left(\frac{\partial \varepsilon_{\bf k}}{\partial k_i}\right)^2 f_{\bf k} (1 - f_{\bf k}),
\end{equation}
one finds that at low temperature, if the Fermi surface is small, the momentum range ${\bf k}$ where $f_{\bf k} (1 - f_{\bf k})$ is non-zero is much larger due to the tendency of the band to rapidly change the Fermi surface topology near the Lifshitz transition (compare the red dashed lines in Fig. 4A) than for a usual Fermi surface.

\section{Conclusion}
To summarize, we reported a CW-like $\chi^{er}(T)$ is observed for all kinds of heavily hole doped 122s.
There is an unexpected enhancement of elastoresistance around the Lifshitz transition. We explained it as a small Fermi pocket effect on the nematicity.
We expect our explanation of an alternative contribution to the enhanced elastoresistance other than a nematic QCP will be considered in other systems, in particular for those with small Fermi pockets.

\section*{Conflict of Interest Statement}
The authors declare that the research was conducted in the absence of any commercial or financial relationships that could be construed as a potential conflict of interest.

\section*{Author Contributions}
IM, SA, VG, KK, and CHL prepared the samples. XCH, FC, and MB did the experiments. SS proposed the theoretical model. CH and BB supervised the study. XCH, SS, FC, and CH analysed the data and wrote the manuscript with input from all authors.

\section*{Funding}
This work has been supported by the Deutsche Forschungsgemeinschaft (DFG) through  SFB 1143 (Project No. 247310070), through the Research Projects CA 1931/1-1 (FC) and SA 523/4-1 (SA). S.S. acknowledges funding by the Deutsche Forschungsgemeinschaft via the Emmy Noether Programme ME4844/1-1 (project id 327807255).
This project has received funding from the European Research Council (ERC) under the European Union's Horizon 2020 research and innovation programme (grant agreement No. 647276-MARS-ERC-2014-CoG).

\section*{Acknowledgments}
We would like to thank Anna B\"{o}hmer, Ian Fisher, Suguru Hosoi, R\"{u}diger Klingeler, Christoph Meingast, J\"{o}rg Schmalian, Christoph Wuttke, Paul Wiecki, and Liran Wang for helpful discussions. We would like to thank Christian Blum and Silvia Seiro for their technical support.

\section*{Data Availability Statement}
The original contributions presented in the study are included in the article; further inquiries can be directed to the corresponding authors.

\end{document}